# A feasibility study for predicting optimal radiation therapy dose distributions of prostate cancer patients from patient anatomy using deep learning

Dan Nguyen, Troy Long, Xun Jia, Weiguo Lu, Xuejun Gu, Zohaib Iqbal, Steve Jiang

Medical Artificial Intelligence and Automation Laboratory, Department of Radiation Oncology, University of Texas Southwestern Medical Center, Dallas, TX, 75390, USA

E-mail: Dan.Nguyen@UTSouthwestern.edu

## Abstract

With the advancement of treatment modalities in radiation therapy for cancer patients, outcomes have improved, but at the cost of increased treatment plan complexity and planning time. The accurate prediction of dose distributions would alleviate this issue by guiding clinical plan optimization to save time and maintain high quality plans. We have modified a convolutional deep network model, U-net (originally designed for segmentation purposes), for predicting dose from patient image contours of the planning target volume (PTV) and organs at risk (OAR). We show that, as an example, we are able to accurately predict the dose of intensity-modulated radiation therapy (IMRT) for prostate cancer patients, where the average Dice similarity coefficient is 0.91 when comparing the predicted vs. true isodose volumes between 0% and 100% of the prescription dose. The average value of the absolute differences in [max, mean] dose is found to be under 5% of the prescription dose, specifically for each structure is [1.80%, 1.03%](PTV), [1.94%, 4.22%](Bladder), [1.80%, 0.48%](Body), [3.87%, 1.79%](L Femoral Head), [5.07%, 2.55%](R Femoral Head), and [1.26%, 1.62%](Rectum) of the prescription dose. We thus managed to map a desired radiation dose distribution from a patient's PTV and OAR contours. As an additional advantage, relatively little data was used in the techniques and models described in this paper.

# 1 Introduction

Radiation therapy has been one of the leading treatment methods for cancer patients, and with the advent and advancements of innovative modalities, such as intensity modulated radiation therapy (IMRT)[1-7] and volume modulated arc therapy (VMAT)[8-14], plan quality has drastically improved over the last few decades. However, such a development comes at the cost of treatment planning complexity. While this complexity has given rise to better plan quality, it can be a double-edged sword that increases the planning time and obscures the tighter standards that these new treatment modalities are capable of meeting. This has resulted in greatly increased clinical treatment planning time, where the dosimetrist goes through many iterations to adjust and tune treatment planning parameters, as well as receiving feedback from the physician many times before the plan is approved. Many further developments in treatment planning algorithms have aided in reducing the treatment complexity, such as including dose-volume constraints in a feasibility seeking algorithm[15], creation of many Pareto surface plans for the planner to navigate through[16-18], and many others for performance improvements and usage simplification[19-25]. However, using any of these algorithms still requires intelligent inputs or tweaks from the human planner, such as weight tuning, deciding appropriate DVH constraints or determining appropriate tradeoffs. To reduce the planning complexity even further, the prediction of dose distributions and constraints has become an active field of research, with the goal of creating consistent plans that are informed by the ever-growing body of treatment planning knowledge, as well as guiding clinical plan optimization to save time and to maintain high quality treatment plans across planners of different experiences and skill levels. Figure 1A shows the typical treatment planning workflow with many iterations for the dosimetrist and physician, and Figure 1B shows the workflow with a dose prediction model in place. Overall workflow does not change, but we expect the number of iterations to considerably decrease.

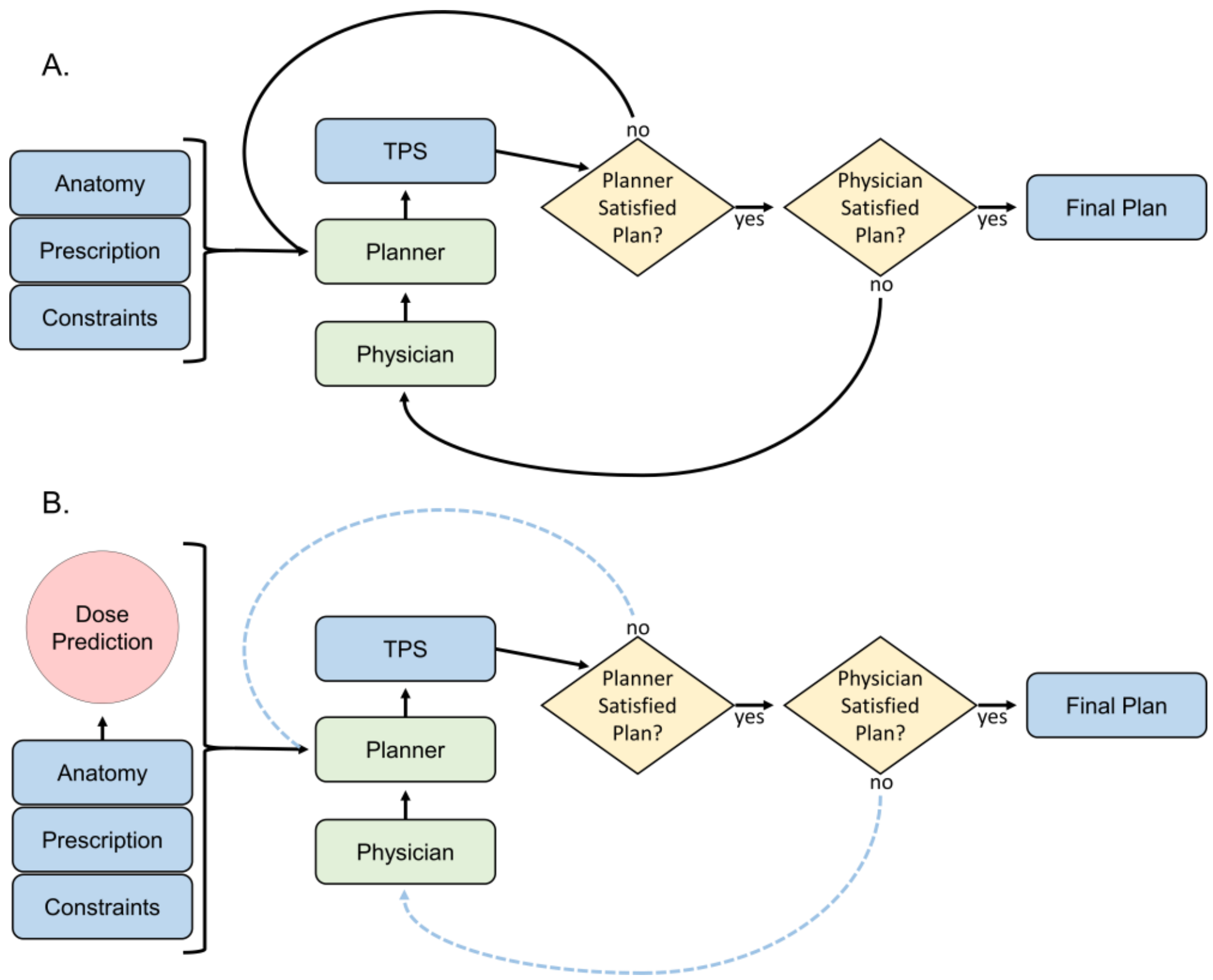

Figure 1: A) Current treatment planning workflow. B) Proposed workflow with AI-based dose prediction. Less iterations denoted as dotted-blue lines. TPS = treatment planning system.

Much of the work for dose prediction in radiotherapy has been revolving around a paradigm known as knowledge-based planning (KBP)[26-38], which has been focused on the prediction of a patient's dose volume histogram (DVH) and dose constraints, using historical patient plans and information. While KBP has seen large successes and advancements that have improved the reliability of its predictions, these methods require the enumeration of parameters/features in order to feed into a model for dose and DVH prediction. Although much time and effort has been spent in selecting handcrafted features—such spatial information of organs at risk (OAR) and planning target volumes (PTV), distance-to-target histograms (DTH), overlapping volume histograms (OVH), structure shapes, number of delivery fields, etc.[31-39]—it is still deliberated as to which features have the greatest impact and what other features would considerably improve the dose prediction. Artificial neural networks have been applied to learn more complex relationships between the handcrafted data[31], but it is still limited by the inherent information present in that data. Another known KBP approach by Good et al.[40], approached the problem by creating a "knowledge database" of 132 prostate treatment plans. A new patient is then matched to one of the knowledge database patient using mutual information as a similarity metric, the database patient's plan parameters are adapted and optimized to the new patient. The success of such a method relies on the size and diversity of its patients, and may possibly be limited when faced with more complex treatment sites, such as head and neck cancer patient.

In the last few years, deep learning has made a quantum leap in the advancement of many areas. One particular area was the progression of convolutional neural network (CNN)[41] architectures for imaging and vision purposes[42-44]. In 2015, fully convolutional networks (FCN)[45] were proposed, and outperformed state-of-the-art techniques of its time at semantic segmentation. Shortly after, more complex models were built around the FCN concept in order to solve some of its shortcomings. One particular architecture that was proposed is a model called U-net[46], which focused on the semantic segmentation on biomedical images. There were three central ideas in the U-net's architecture design: 1) a large number of max pooling operations to allow for the convolution filters to find global, non-local features, 2) transposed convolution operations—also known as deconvolution[47] or up-convolution[46]—to return the image to its original size, and 3) copying the maps from the first half of the U-net in order to preserve the lower-level, local features. While inserting some domain knowledge into the problem may be helpful due to a limited amount of data, we look towards deep learning to reduce our dependence on handcrafted features, and allow the deep network to learn its own features for prediction. Even though the U-net and other FCN architectures were designed for the task of image segmentation, we hypothesize that, with some innovative modifications, the U-net architecture will be able to accurately predict a voxel-level dose distribution simply from patient contours, by learning to abstract its own high-level local and broad features. Our motivation is two-fold: 1) (short term motivation) to provide guidance for the dosimetrist during clinical plan optimization in order to improve the plan quality and uniformity and, to reduce the total planning time by decreasing the number of iterations the dosimetrist has to go through with the physician and treatment planning optimization, and 2) (long term motivation) to eventually develop an artificial intelligent treatment planning tool, capable of creating entire clinically acceptable plans.

# 2 Methods

## 2.1 U-net architecture for dose prediction

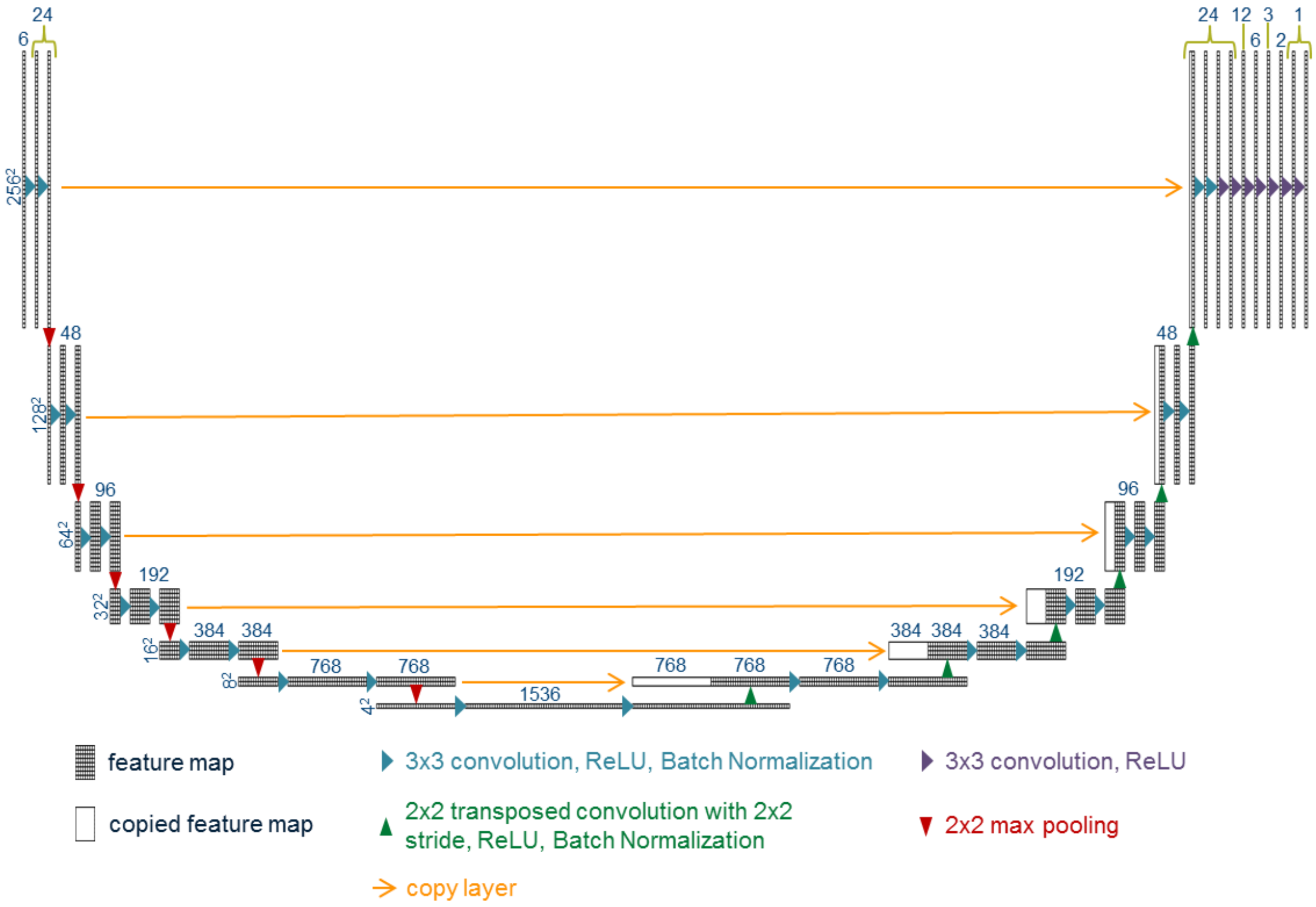

Figure 2: Schematic of an example U-net architecture with additional CNN layers used for dose prediction. The numbers above the boxes represent the number of features for each map, while the numbers to the left of each hierarchy in the U-net represents the size of each 2D feature.

As shown in Figure 2, we constructed a seven-level hierarchy U-net, with some innovative modifications made on the original design achieve the goal of contour-to-dose mapping. The input starts with 6 channels of 256 x 256 pixel images. Specifics of the input data is outlined in Section 2.2. The choice for 7 levels with 6 max pooling operations was made to reduce the feature size from 256 x 256 pixels down to 4 x 4 pixels, allowing for the 3 x 3 convolution operation to connect the center of the tumor to the edge of the body for all of the patient cases. Zero padding was added to the convolution process so that the feature size is maintained. Seven CNN layers, denoted with the purple arrows in Figure 2, were added after the U-net in order to smoothly reduce the number of filters to one, allowing for high precision prediction. Batch normalization[48] (BN) was added after the convolution and rectified linear unit (ReLU) operations in the U-net, which allows for a more equal updating of the weights throughout the U-net, leading to faster convergence. It should be noted that the original BN publication suggests performing the normalization process before the non-linearity operation, but we had found better performance using normalization after the ReLU operation—the validation's mean squared error after 10 epochs was 0.3528 for using BN before ReLU and 0.0141 for using BN after ReLU.

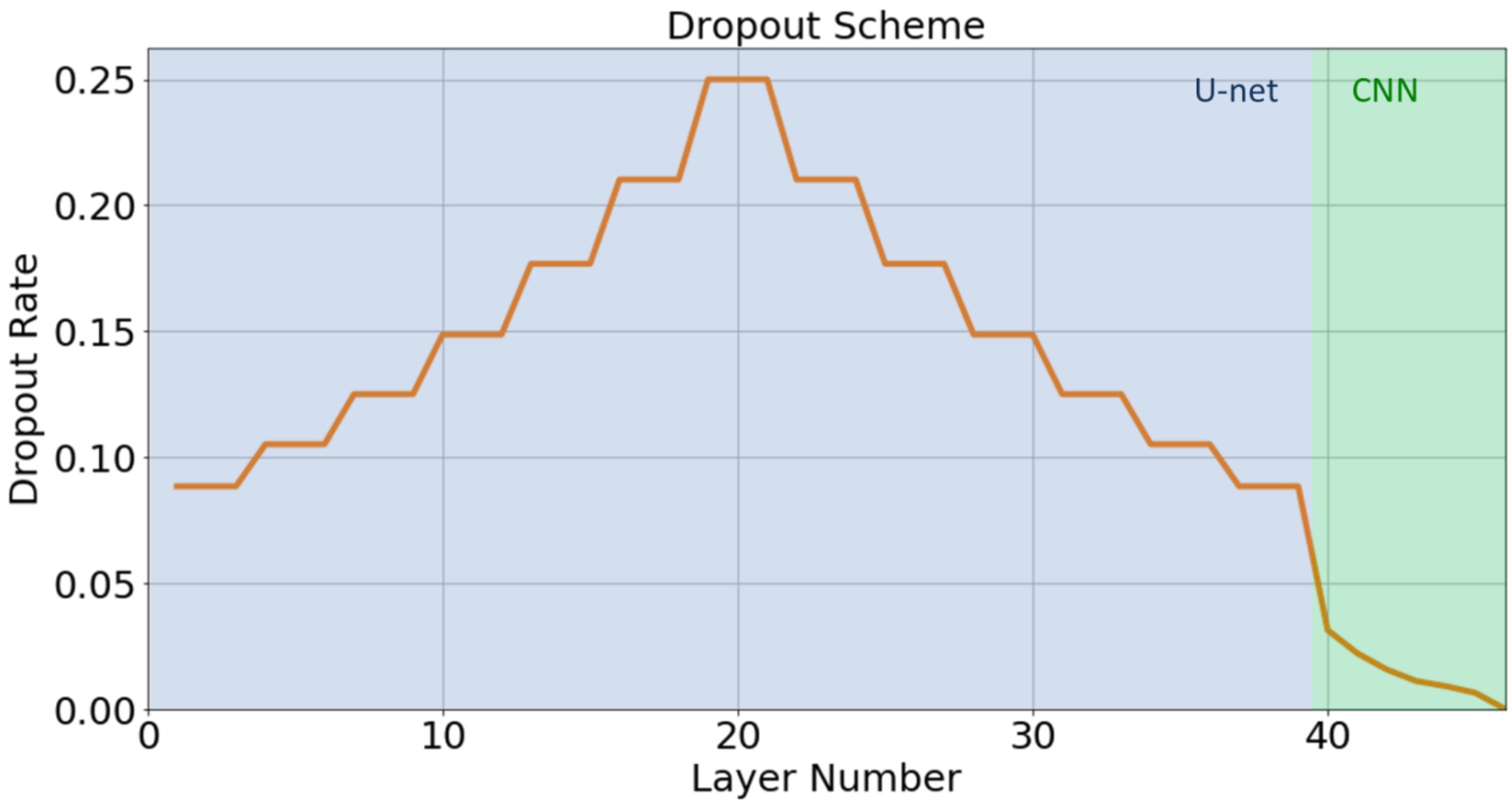

Figure 3: Dropout scheme implemented for the U-net and CNN layers

To prevent the model from over-fitting, dropout[49] regularization was implemented according to the scheme shown in Figure 3, which is represented by the equation: $dropout_{rate} = rate_{max} \times \left(\frac{current\ number\ of\ filters}{\max number\ of\ filters}\right)^{1/n}$. For our setup, we chose $rate_{max} = 0.25$ and the $max\ number\ of\ filters = 1536$. We chose $n = 4$ for the U-net layers, and $n = 2$ for the added CNN layers. The choice for the dropout parameters was determined empirically, until the gap between the validation loss and training loss did not tend to increase during training.

The Adam algorithm[50] was chosen as the optimizer to minimize the loss function. We used a learning rate of $1 \times 10^{-4}$, and the default Adam parameters $\beta_1 = 0.9$, $\beta_2 = 0.999$, and $decay = 0$. In total, the network consisted of 46 layers. The deep network architecture was implemented in Keras[51] with Tensorflow[52] as the backend.

## 2.2 Training and Evaluation

To test the feasibility of this model, treatment plans of 88 clinical coplanar IMRT prostate patients, each planned with 7 IMRT fields at 15 megavolts (MV), were used. The 7 IMRT beam angles were similar across the 88 patients. Each patient had 6 contours: planning target volume (PTV), bladder, body, left femoral head, right femoral head, and rectum. The volume dimensions were reduced to 256 x 256 x 64 voxels, with resolutions of 2 x 2 x 2.5 $mm^3$. For training, all patient doses were normalized such that the mean dose delivered to the PTV was equal to 1.

The U-net model was trained on single slices of the patient. As input, the 6 contours were each treated as their own channel in the image (analogous to how RGB images are treated as 3 separate channels in an image). The output is the U-net's prediction of the dose for that patient slice. The loss function was chosen to be the mean squared error between the predicted dose and the true dose delivered to the patient.

Since the central slices containing the PTV were far more important than the edge slices for dose prediction, we implemented a Gaussian sampling scheme—the center slice would more likely be chosen when the training function queried for another batch of random samples. The distance from the center slice to the edge slice was chosen to equal 3 standard deviations for the Gaussian sampling.

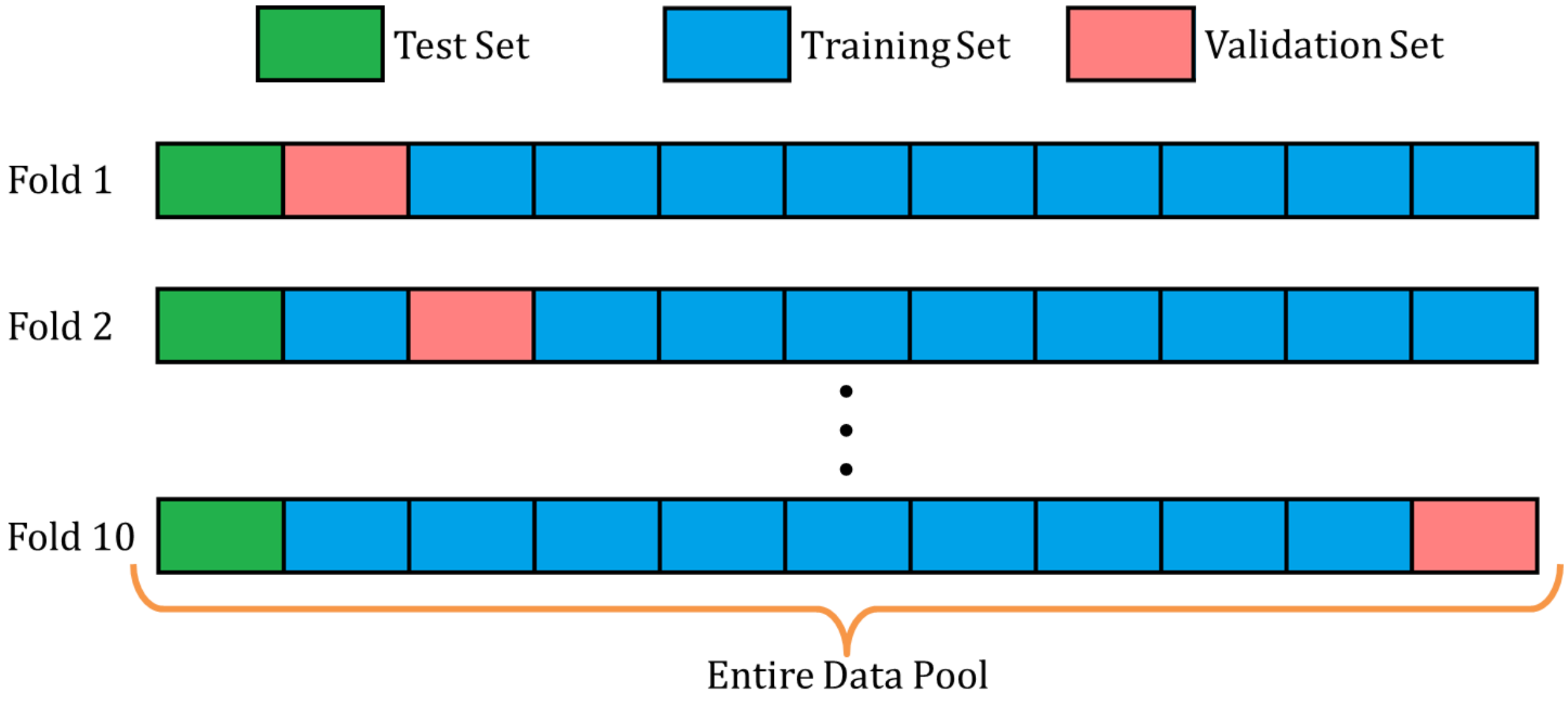

Figure 4: Schematic for 10-fold cross-validation. A test set is held out from the cross validation procedure, and is used to test the best performance model.

To assess the overall performance of the model, 8 patients were selected as a test set, and then 10-fold cross-validation procedure was performed on the remaining 80 patients, as shown in Figure 4. Each of the 10 folds divides the remaining 80 patients into 72 training patients and 8 validation patients. Ten separate U-net models are initialized, trained, and validated on a unique training and validation combination. Each fold produces a model that can predict a dose distribution from contours. From these 10 trained models, we then take the best performance model, based on its validation loss, and evaluate this model on the test set.

For the remainder of the manuscript, some common notation will be used. $D\#$ is the dose that #% of the volume of a structure of interest is at least receiving. $V_{ROI}$ is the volume of the region of interest. For example, $D95$ is the dose that 95% of the volume of the structure of interest is at least receiving. $V_{PTV}$ is the volume of the PTV and $V_{\#\%Iso}$ is the volume of the #% isodose region. Isodose volumes are binary masks defined as 1 if the voxel contains a dose value above some threshold and 0 otherwise. The #% is the threshold for the isodose volume calculation and represents a percent of the prescription dose.

To equally compare across the patients, all plans were normalized such that 95% of the PTV volume was receiving the prescription dose ($D95$). This normalization is applied to both the dose in the test set and to the dose prediction of the network, and was done by multiplying the dose with ratio $\frac{prescription\ dose}{current\ dose\ delivered\ to\ 95\%\ of\ PTV}$. It should be noted that this is normalized differently than for training the model, which had normalized the plans by PTV mean dose. Normalizing by PTV mean dose creates a uniform dataset which is more likely to be stable for training, but plans normalized by D95 have more clinical relevance and value for assessment. All dose statistics will also be reported relative to the prescription dose (i.e. the prescription dose is set to 1). As evaluation criteria, Dice similarity coefficients $\left(\frac{2(A \cap B)}{A+B}\right)$ of isodose volumes, structure mean and max doses, PTV $D98$, $D99$, $D_{max}$, PTV homogeneity $\left(\frac{D2-D98}{D50}\right)$, van't Riet conformation number[53] $\left(\frac{(V_{PTV} \cap V_{100\%Iso})^2}{V_{PTV} \times V_{100\%Iso}}\right)$, and the dose spillage $R50$ $\left(\frac{V_{50\%Iso}}{V_{PTV}}\right)$, were evaluated.

Five NVIDIA Tesla K80 dual-GPU graphics cards (10 GPU chips total) were used in this study. One GPU was used for training each fold of the 10-fold cross-validation. Training batch size was chosen to be 24 slices. The datasets generated during and/or analyzed during the current study are not publicly available due to sensitive medical information but are available from the corresponding author on reasonable request. Usage of the patient data has been approved by the UT Southwestern Protocol Review and Monitoring Committee (PRMC) and the Institutional Review Board (IRB). All patient data has been fully anonymized, and all methods were performed in accordance with the relevant guidelines and regulations outlined by the institution. Since gathered patient data was retrospective and did not directly involve the human participants during the study, informed consent is not applicable to this study.

## 3 Results

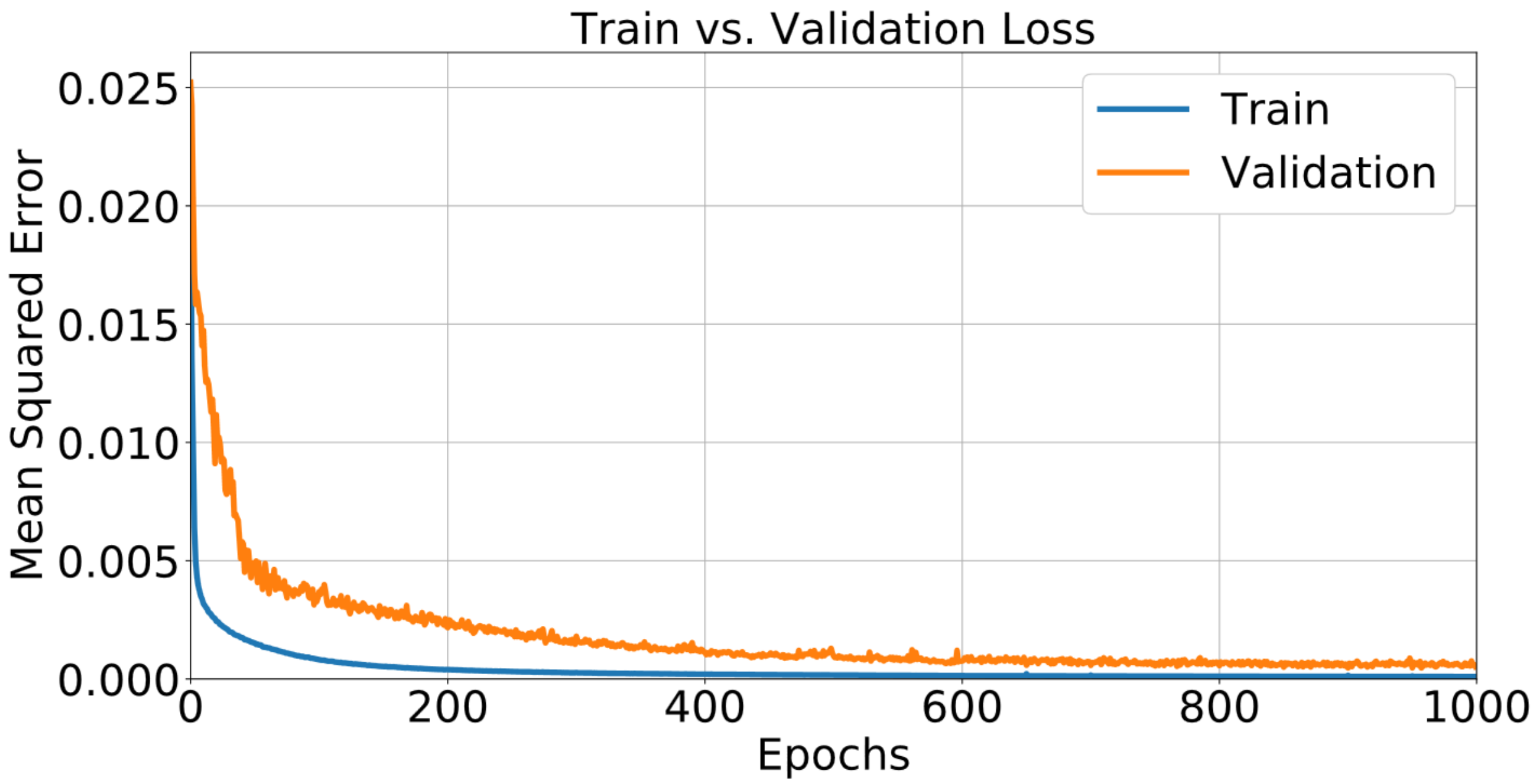

Figure 5: Plot of train vs. validation loss as a function of epochs from one of the folds.

In total, models from all folds trained for 1000 epochs each, which took approximately 6 days on the 10 GPUs. A plot of training and validation loss from one of the folds is shown in Figure 5 as an example. The final average loss ± standard deviation between all the folds is $(1.02 \pm 0.05) \times 10^{-4}$ (training loss) and $(6.26 \pm 1.34) \times 10^{-4}$ (validation loss). Of the 10 folds, the model from the 5th fold performed the best with the lowest validation loss of $4.47 \times 10^{-4}$. This model was used to evaluate the dosimetric performance on the test set of patients.

| | Average Absolute Dose Difference $\left\|\frac{D_{True}-D_{Prediction}}{D_{Prescription}}\right\| \times 100$ | | | |
|---|---|---|---|---|
| | mean value ± standard deviation | | | |
| | Cross-Validation Results | | Test Results | |
| | $D_{max}$ | $D_{mean}$ | $D_{max}$ | $D_{mean}$ |
| PTV | 1.41 ± 1.13 | 0.77 ± 0.58 | 1.80 ± 1.09 | 1.03 ± 0.62 |
| Bladder | 1.38 ± 1.17 | 2.38 ± 2.26 | 1.94 ± 1.31 | 4.22 ± 3.63 |
| Body | 1.45 ± 1.21 | 0.86 ± 0.42 | 1.80 ± 1.09 | 0.48 ± 0.35 |
| L Fem Head | 2.46 ± 2.56 | 1.16 ± 0.74 | 3.87 ± 3.26 | 1.79 ± 1.58 |
| R Fem Head | 2.42 ± 2.45 | 1.17 ± 0.88 | 5.07 ± 4.99 | 2.55 ± 2.38 |
| Rectum | 1.34 ± 1.02 | 1.39 ± 1.03 | 1.26 ± 0.62 | 1.62 ± 1.07 |

Table 1: Average differences in mean and max dose with standard deviations.

| | PTV Statistics, van't Riet Conformation Number, and Dose Spillage | | | | | |
|---|---|---|---|---|---|---|
| | mean value ± standard deviation | | | | | |
| | Cross-Validation Results | | | Test Results | | |
| | True Values | Pred Values | True - Pred | True Values | Pred Values | True - Pred |
| PTV D98 | 0.98 ± 0.01 | 0.98 ± 0.01 | -0.00 ± 0.01 | 0.98 ± 0.01 | 0.98 ± 0.01 | 0.00 ± 0.00 |
| PTV D99 | 0.97 ± 0.01 | 0.97 ± 0.04 | 0.00 ± 0.04 | 0.96 ± 0.01 | 0.97 ± 0.01 | 0.00 ± 0.01 |
| PTV $D_{max}$ | 1.08 ± 0.02 | 1.08 ± 0.02 | 0.01 ± 0.02 | 1.08 ± 0.01 | 1.07 ± 0.02 | 0.01 ± 0.02 |
| PTV Homogeneity | 0.09 ± 0.02 | 0.08 ± 0.03 | 0.01 ± 0.02 | 0.09 ± 0.01 | 0.07 ± 0.02 | 0.01 ± 0.02 |
| van't Riet Conformation Number | 0.88 ± 0.08 | 0.92 ± 0.04 | -0.04 ± 0.05 | 0.91 ± 0.02 | 0.90 ± 0.03 | 0.00 ± 0.02 |
| R50 | 4.45 ± 1.23 | 4.10 ± 1.14 | 0.35 ± 0.23 | 4.00 ± 0.37 | 3.98 ± 0.32 | 0.02 ± 0.21 |

Table 2: True and predicted values for PTV statistics, homogeneity, van't Riet conformation number, and the high dose spillage, R50.

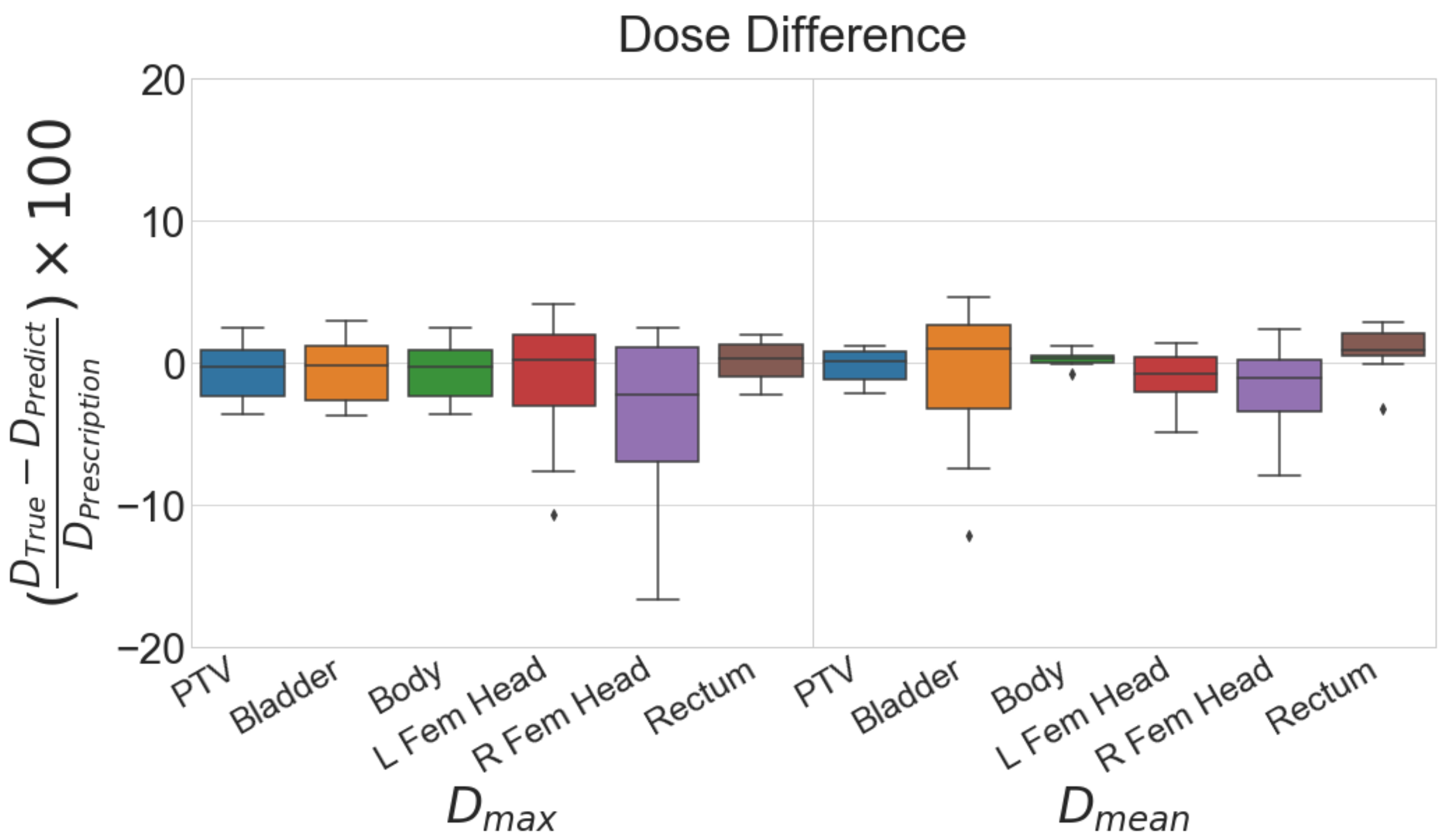

Figure 6: Box plots showing the dose difference statistics for the 8 test patients.

A box plot of max and mean dose differences (True – Prediction) for the PTV and OARs for the test patient cases are shown in Figure 6. On average, the U-net model is biased to slightly over-predict the mean and max doses. A full list of average absolute differences for both the cross validation and test data can be found in Table 1. Overall, the cross validation error is slightly less than the test error. For the test data, the PTV, body and rectum maintain a prediction accuracy of within 3% error. The bladder has a low max dose error of 1.9% but a larger error in the mean dose of 4.2%. The femoral heads have higher max dose errors but reduced mean dose errors of under 3%. Overall, the model is capable of accurately predicting $D_{max}$ and $D_{mean}$ within 5.1% of the prescription dose. In addition all of the PTV related dosimetric statistics, dose conformity, and the dose spillage, $R50$, are very well predicted by the network as shown in Table 2. The PTV coverage, PTV $D_{max}$, conformation number, and R50 have less than 1% error (calculated as $\left|\frac{True-Predicted}{True}\right| * 100$).

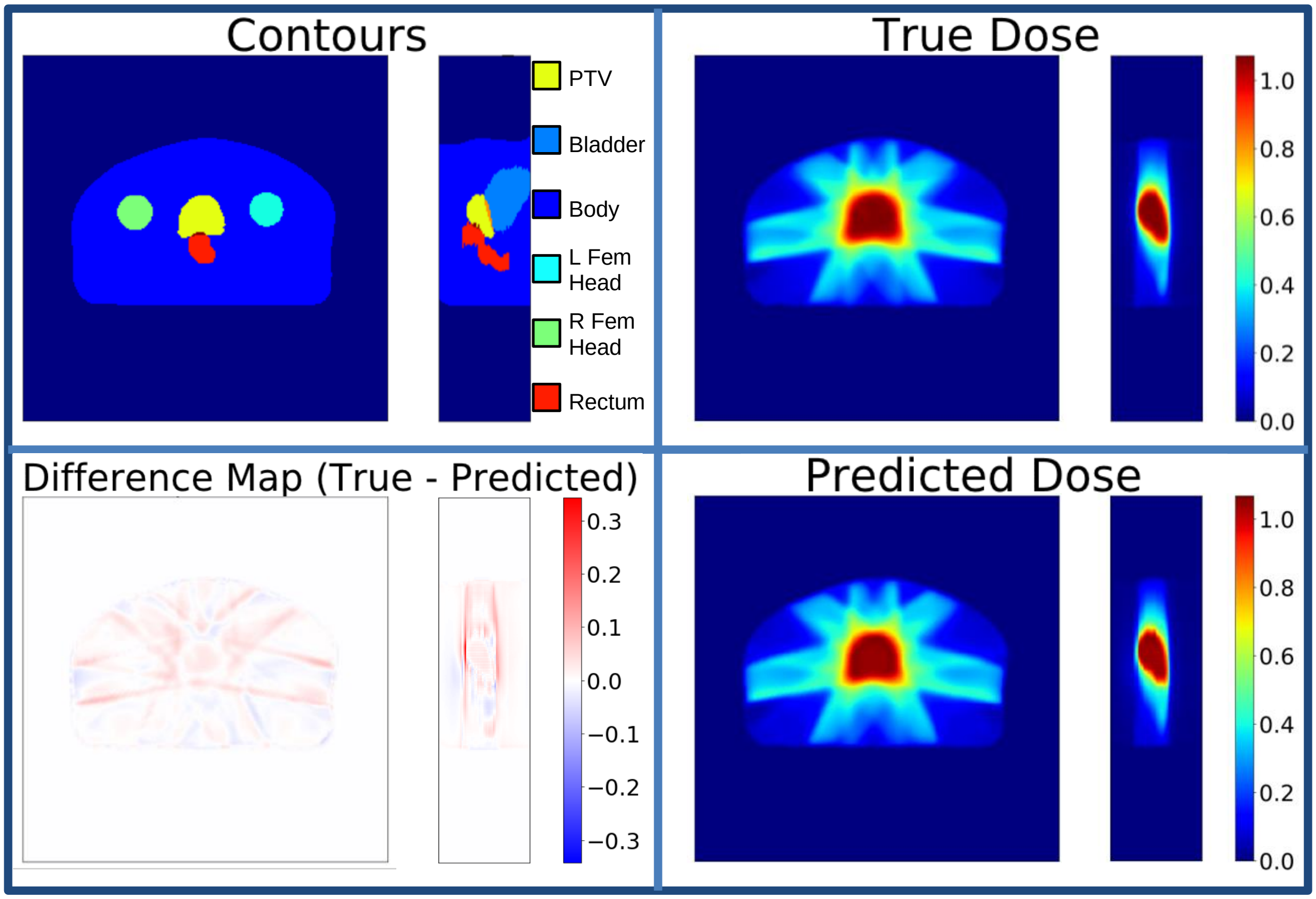

Figure 7: Contours of the planning target volume (PTV) and organs at risk (OAR), true dose wash, predicted dose wash, and difference map of an example patient.

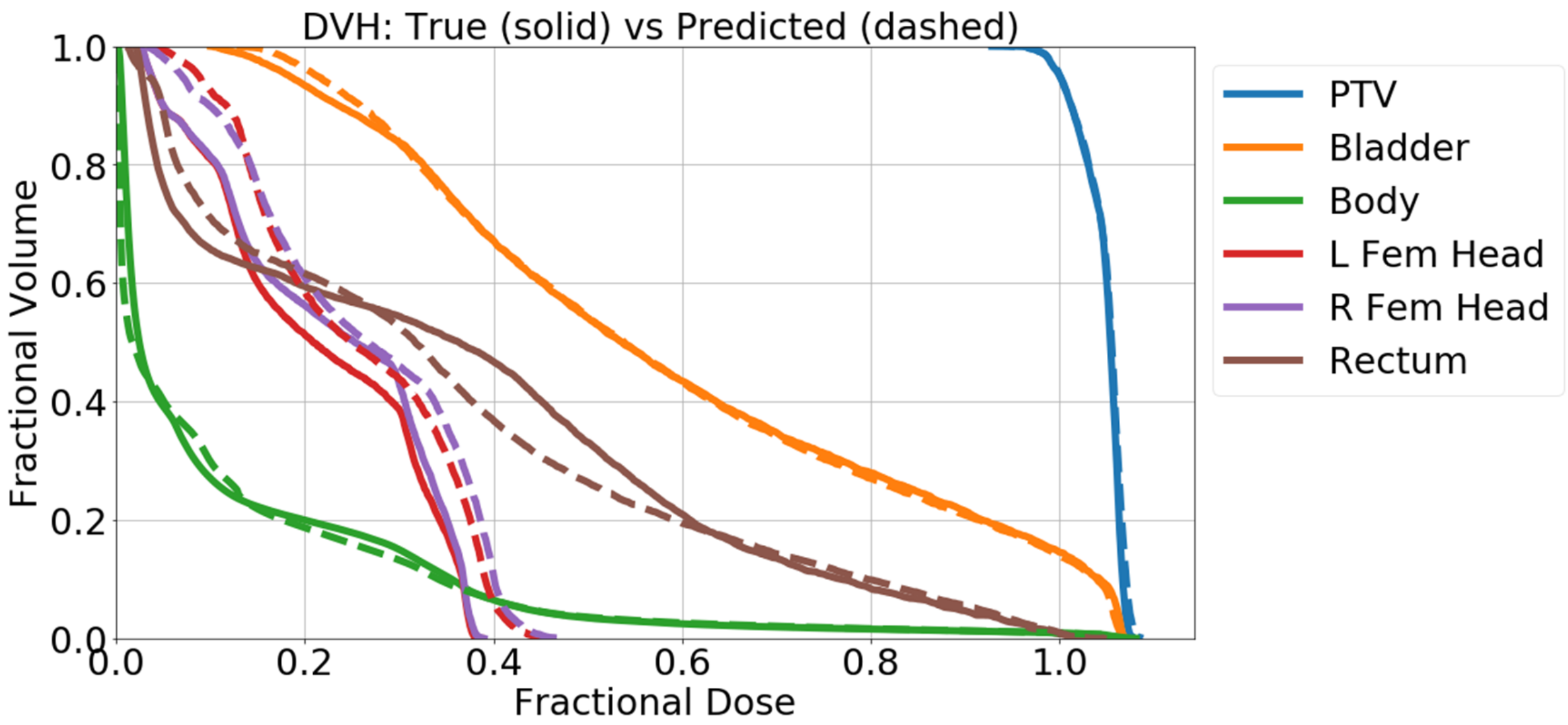

Figure 8: Example of typical dose volume histogram (DVH) comparing true dose and predicted dose for one patient.

As a typical prediction example from the U-net model, Figure 7 shows the input contours, true and predicted dose washes, and a difference map of the two doses for one patient. On average, the dose difference inside the body was less than 1% of the prescription dose, shown in Table 1. Figure 8 shows the DVH of one of the example test patients. Visually on the DVH, one can see that the U-net tends to predict a similar PTV dose coverage with minimal errors in the dose prediction to the OARs.

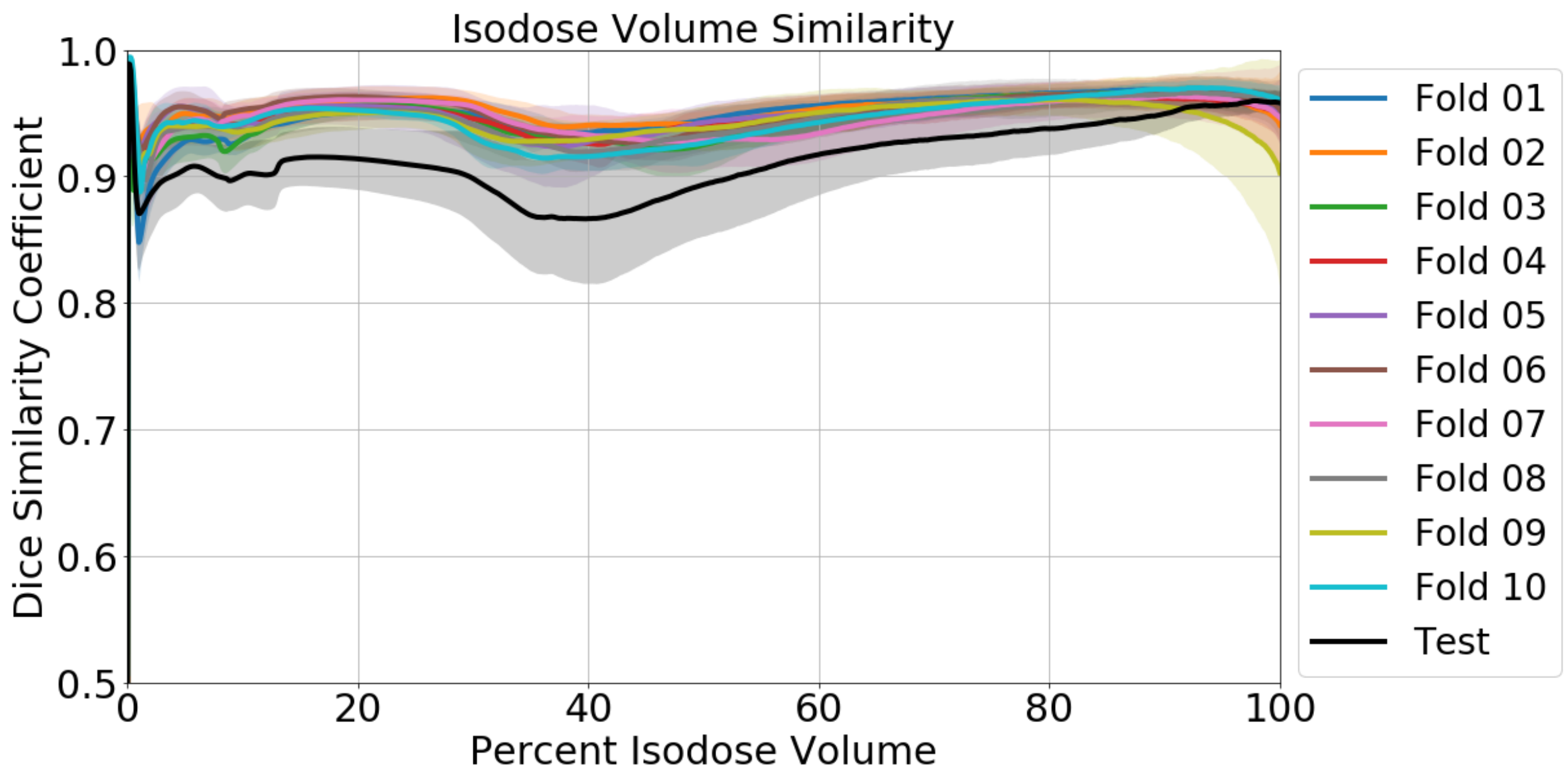

Figure 9: Dice similarity coefficients, $\frac{2(A \cap B)}{A+B}$, comparing isodose volumes between the true dose and predicted dose, ranging from the 0% isodose volume to the 100% isodose volume. The error in the graph represents 1 standard deviation.

The plot of Dice similarity coefficients of isodoses is shown in Figure 9. Dice similarity coefficients range from 0 to 1, where 1 is considered a perfect match. The average Dice similarity coefficient for the test data is 0.91 and for the cross-validation data is 0.95, a 4% difference. The isodose volume similarity expresses slight decreases in the Dice coefficient near the 40% isodose volume. The loss in predictability at 40% is associated to the complicated details in the dose distribution along the beam paths in the normal tissue, which is generated during the fluence map optimization process.

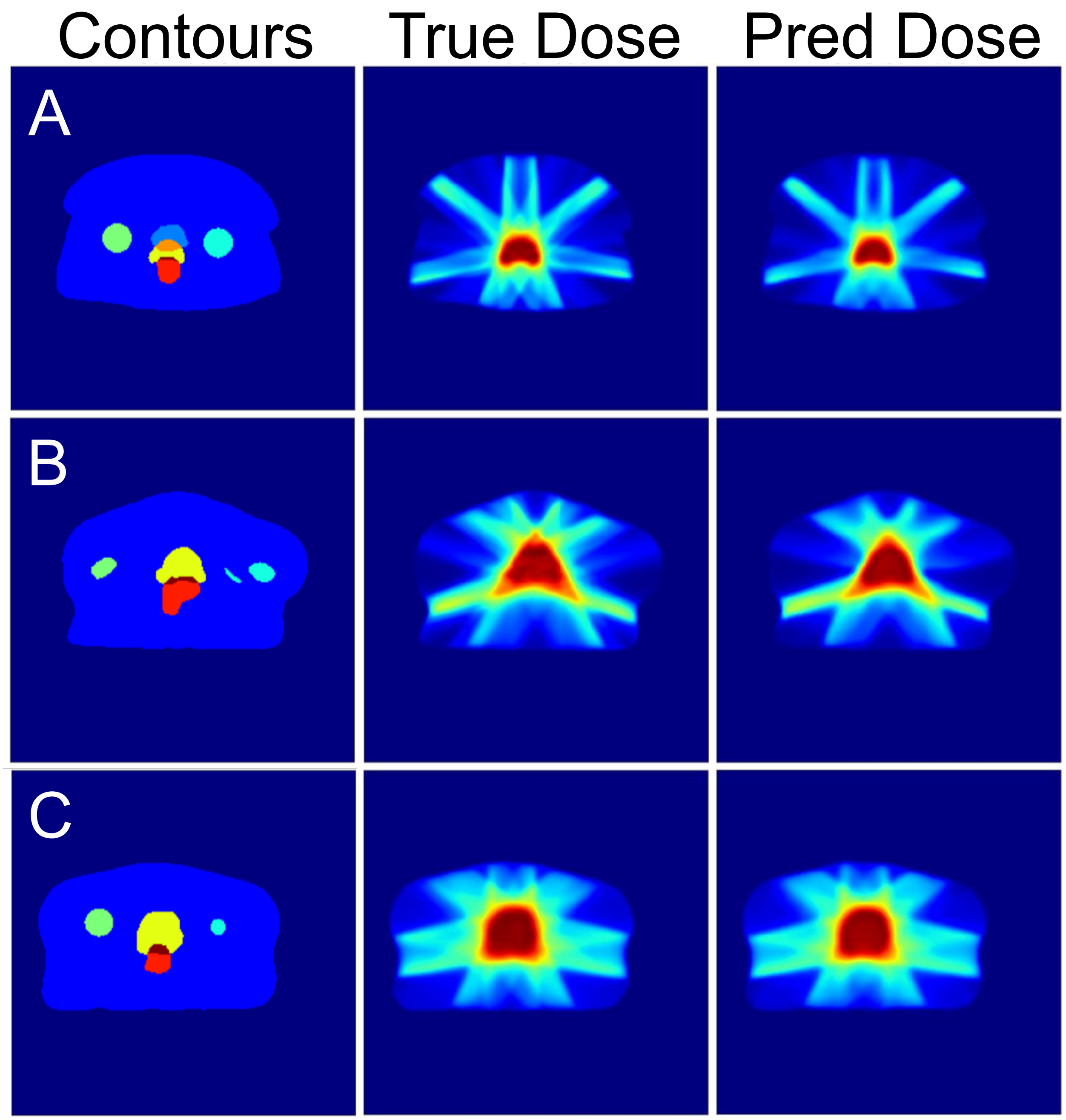

Figure 10: Example dose predictions from the U-net model on several patients with vastly different geometries.

Figure 10 shows some examples of dose prediction from the U-net on patients that have very diverse geometries. It can be visually seen that the U-net has learned to shape the dose based on the PTV and OARs sizes, locations, and shapes. The finer details of the dose distributions further away from the PTV have been predicted by the deep network model with relative high accuracy.

# 4 Discussion

To our knowledge, this is the first fully 3D dose distribution prediction for prostate IMRT plans, thus making direct comparison to existing models difficult. The latest study by Shiraishi and Moore[31] on knowledge based planning did investigate 3D dose prediction, but for prostate patients treated with VMAT. In addition, another cutting edge study by McIntosh and Purdie[54] investigated 3D dose prediction using atlas regression forests. Because of the differing patient data base and treatment modalities/protocols, the results cannot be directly compared. It should be noted that Shiraishi and Moore's average prediction error was less than 8% using their method on their patients, and McIntosh and Purdie's study found the average Dice coefficient to be 0.88 (range is from 0.82 to 0.93). Therefore, our impression is that our predictive model is at least within the same ballpark as the cutting edge methods by these authors.

The 88 clinical prostate patients acquired in this study used a similar set of 7 beam angles and criteria for treatment, giving rise to some uniformity to the data that made it ideal as a test bed to investigate the feasibility for dose prediction using a deep learning model. However, the current model architecture and data leave the U-net with several limitations. First, the model has currently learned to only predict the dose coming from approximately the same orientations, and may not be able to account for more intricate beam geometries. Secondly, the current model is unable to account for any physician preferences for predicting the dose, limiting the level of treatment personalization for the patient. For example, the model is unable to create a rectum-sparing plan or a bladder-sparing plan, at the will of the physician, for the same patient geometry. Furthermore, while training slice-by-slice had proven successful for coplanar cases, this method may not perform satisfactorily when performing dose prediction for non-coplanar plans. The deep network may have to understand the patient geometry in 3D if it were to start accounting for non-coplanar beam dose. Nevertheless, because the clinical prostate IMRT protocol is standardized, the current dose prediction model from this study can still be employed as a clinical guidance tool, where final tradeoff decisions will still be made by the physician and dosimetrist. By utilizing this model, the physician can immediately view the dose prediction and then convey how they desire for the plan to be changed to the dosimetrist. By already having a tangible plan to view, the dosimetrist can more readily apply the changes to make an acceptable plan earlier and ultimately reduce the total planning time.

We plan to extend this study by building a deep learning model for learning dose predictions that is capable of handling a more diverse selection of non-coplanar beam orientations. We will investigate the extension of U-nets into the volumetric domain using V-nets[55], in order to tackle dose prediction for non-coplanar radiotherapy plans, and add in dose constraint parameters into the model input to allow the prediction of dose based on the physician's prescription, not only patient's geometry. Furthermore, we will examine the addition of CT data and its effect on prediction accuracy. We expect the addition of such information to the model will greatly improve the prediction accuracy, and will investigate the impact of adding these types of information.

## 5 Conclusion

We have developed a novel application of the fully convolutional deep network model, U-net, for dose prediction. The model is able to take a prostate patient's contours and then predict a dose distribution by abstracting the contours into local and global features. Using our implementation of U-net we are able to accurately predict the dose of a patient, with average mean and max dose differences of all structures within 5.1% of the prescription dose. Isodose similarity evaluation reveals that the predicted dose isodose volumes match the true isodose volumes with the average Dice coefficient of 0.91. We plan to continue improving the model, by adding in dose prediction for non-coplanar beam arrangements and accounting for physician preference. The immediate application of the dose prediction model is to guide clinical plan optimization to reduce treatment planning time and to maintain high quality plans. The long-term objective is to incorporate the learning dose prediction model into an artificially intelligent radiation therapy treatment planner.

## 6 Author Contributions

Dan Nguyen and Steve Jiang conceived the experiment. Dan Nguyen coded the neural network, conducted the experiment, analyzed the results, generated all the figures, and composed the manuscript. Troy Long contributed several key ideas to the experimental approach and neural network training scheme. Xun Jia, Weiguo Lu, Xuejun Gu, and Zohaib Iqbal equally contributed ideas regarding the U-net architecture and cross-validation procedure with a hold-out test set. Steve Jiang oversaw the overall project direction. All authors reviewed the manuscript.

## 7 Acknowledgements

This study was supported by the Cancer Prevention & Research Institute of Texas (CPRIT) IIRA RP150485.

## 8 Competing Interests

The authors declare no competing interests.

## 9 Protection of Human Subjects

Usage of the patient data has been approved by the UT Southwestern Protocol Review and Monitoring Committee (PRMC) and the Institutional Review Board (IRB). All patient data has been fully anonymized.